\documentclass[12pt,preprint]{aastex}
\usepackage{amsmath}

\cpright{none}{2022}

\shortauthors{Wang}

\shorttitle{Undetected Minority-polarity Flux and Coronal Heating}

\begin{document}

\title{Undetected Minority-polarity Flux as the Missing Link in Coronal Heating}
\author{Y.-M. Wang}
\affil{Space Science Division, Naval Research Laboratory, Washington, DC 20375, USA}
\email{yi.wang@nrl.navy.mil}

\begin{abstract}
During the last few decades, the most widely favored models for 
coronal heating have involved the in situ dissipation of energy, 
with footpoint shuffling giving rise to multiple current sheets 
(the ``nanoflare'' model) or to Alfv{\'e}n waves that leak into the corona 
and undergo dissipative interactions (the wave heating scenario).  
As has been recognized earlier, observations suggest instead 
that the energy deposition is concentrated at very low heights, 
with the coronal loops being filled with hot, dense material from below, 
which accounts for their overdensities and flat temperature profiles.  
While an obvious mechanism for footpoint heating would be reconnection 
with small-scale fields, this possibility seems to have been widely 
ignored because magnetograms show almost no minority-polarity flux 
inside active region (AR) plages.  Here, we present further examples 
to support our earlier conclusions (1) that magnetograms greatly 
underrepresent the amount of minority-polarity flux inside plages and 
``unipolar'' network, and (2) that small loops are a major constituent of 
\ion{Fe}{9} 17.1 nm moss.  On the assumption that the emergence or 
churning rate of small-scale flux is the same inside plages as in 
mixed-polarity regions of the quiet Sun, we estimate the energy flux density 
associated with reconnection with the plage fields to be on the order 
of 10$^7$ erg cm$^{-2}$ s$^{-1}$, sufficient to heat the AR corona.
\end{abstract}


\section{INTRODUCTION}
Coronal heating models are generally of two kinds: those that invoke 
magnetohydrodynamic (MHD) waves and their dissipation, and those 
that rely more directly on magnetic reconnection or the formation 
of current sheets (see, e.g., Parnell and De Moortel, 2012, and references 
therein).  In both cases, footpoint motions are usually taken to be 
the driving force and granular convection the basic energy source.  
Although it is possible (and even likely) that both wave dissipation and 
reconnection are involved in coronal heating, there is no consensus as 
to which process is the dominant one.

Alfv{\'e}n waves generated at the photosphere are strongly reflected 
due to the steep falloff of the density and increase in the Alfv{\'e}n speed 
with height, and only a very small fraction of the wave energy leaks 
through the transition region into the corona (see, e.g., Cranmer and 
Ballegooijen, 2005).  Given the presence of a sufficiently large 
wave flux density in the corona, the mechanisms by which the 
Alfv{\'e}n waves are dissipated (resonant excitation, phase mixing, 
turbulent interactions between outgoing and reflected waves) continue 
to be debated (see, e.g., De Groof and Goossens, 2002; Howson et al., 2019; 
van Ballegooijen et al., 2014).

Heating by reconnection has been well observed in flares and jets, 
on scales ranging from X-ray and extreme-ultraviolet (EUV) bright points 
(and the ``campfires'' discovered by {\it Solar Orbiter}: 
Berghmans et al., 2021) to active regions (ARs).  In the well-known 
nanoflare model of Parker (1988), random footpoint motions spontaneously 
give rise to tangential discontinuities/current sheets in the 
coronal field, leading to multiple reconnection events (``nanoflares'') 
and ohmic dissipation (for a recent discussion of this model, 
see Pontin and Hornig, 2020).  One possible difficulty with 
this scenario is that the upper parts of coronal loops generally appear 
smooth and featureless, seemingly inconsistent with the large departures 
from a potential field that would be associated with the predicted 
magnetic braiding (van Ballegooijen et al., 2011).  As emphasized by 
Aschwanden et al. (2007), it is at the footpoints of coronal loops 
that topologically complex structures are observed.  They also argued 
that the high densities in AR loops and the absence of large 
temperature gradients along them can only be explained if the corona 
is not heated locally but from below, via chromospheric evaporation.

It has previously been suggested that small bipoles or ephemeral regions 
(ERs) are the main source of coronal heating in the quiet Sun 
(Schrijver et al., 1998) and in coronal holes (Parker, 1991).  
However, a strong argument against reconnection with small-scale 
fields playing a major role in heating coronal holes and the AR corona 
was provided by the statistical study of ERs by Hagenaar et al. (2008, 2010).  
From an examination of {\it Michelson Doppler Interferometer} (MDI) 
magnetograms recorded during 2000--2005, they concluded that the 
rate of ER emergence is at least a factor of 3 smaller inside 
predominantly unipolar areas than in regions where the polarities 
are more mixed.  This reduction is too large to be attributed to the 
shorter cancellation timescales of minority-polarity flux in a 
strongly unipolar background.

In their analysis of continual intermittent outflow observed in AR loops 
with the {\it Transition Region and Coronal Explorer} (TRACE), 
Winebarger et al. (2001) conjectured that the flows were driven by 
small-scale reconnection events, but found no evidence for underlying 
flux cancellation in MDI magnetograms.  Similarly, Aschwanden et al. (2007) 
mentioned as a ``caveat'' against the chromospheric evaporation scenario 
``the nondetection of the initial heating agent (small-scale reconnection 
events in the transition region)''.  More recently, in their observational 
study of the relation between magnetic structures and signatures of heating 
in a plage region, Anan et al. (2021) found that the fraction of 
opposite-polarity fields was too small to support reconnection mechanisms, 
except possibly in the ``strong guide-field regime.''

However, we have recently come to question the reliability of 
present-day magnetograms in detecting minority-polarity flux in the 
presence of a strongly dominant polarity.  In an analysis of coronal 
EUV plumes employing data from the {\it Atmospheric Imaging Assembly} (AIA) 
and {\it Helioseismic and Magnetic Imager} (HMI) on the 
{\it Solar Dynamics Observatory} (SDO), Wang et al. (2016) found that 
strong plume emission sometimes occurred above network concentrations 
that were purely unipolar according to the HMI magnetograms, but 
the corresponding AIA images clearly showed small, looplike structures 
in the cores of the plumes.  In subsequent studies focusing on 
unipolar plage areas, Wang (2016) and Wang et al. (2019) identified 
large numbers of looplike features embedded within the plages, 
some having the inverted-Y topology characteristic of jets.  
The footpoint structures had horizontal extents of 
$\sim$2$^{\prime\prime}$--7$^{\prime\prime}$ ($\sim$1.5--5 Mm), 
greatly exceeding the HMI pixel size of 0.5$^{\prime\prime}$; the absence 
of corresponding minority-polarity signatures thus appears to be related to 
the magnetograph's sensitivity to weak signals when one polarity is heavily 
dominant, and not simply the result of inadequate spatial resolution.

Analyzing $\sim$0.1$^{\prime\prime}$-resolution measurements of an 
emerging AR from the {\it SUNRISE} {\it Imaging Magnetograph Experiment} 
(IMaX), Chitta et al. (2017) found that the footpoints of coronal loops 
were often located near minority-polarity flux that was invisible 
or barely visible in the corresponding lower-resolution HMI magnetograms.  
However, the minority-polarity elements detected with IMaX were confined 
to the edges of the strong plage areas, whereas the small looplike features 
identified in Wang (2016) and Wang et al. (2016, 2019) were located 
well inside the supposedly unipolar plages and network concentrations, 
as well as at their peripheries.

In this paper, we describe more examples of looplike features and compact 
brightenings at the footpoints of AR loops embedded in ``unipolar'' 
plages, and discuss the nature and origin of the small loops and their 
relationship to AR moss.  We then argue that they provide an energy flux 
sufficient to heat the AR corona.

\section{The Energization of AR Loops by Footpoint Reconnection: SDO Observations}
The AIA instrument records full-disk images in seven EUV channels, 
with 0.6$^{\prime\prime}$ pixels and 12 s cadence, as well as in 
two UV and a white-light channel.  HMI provides line-of-sight magnetograms 
every 45 s, with a noise level of $\sim$10 G.  For this study, 
we employ AIA images taken in three passbands: 17.1 nm, dominated by 
\ion{Fe}{9} ($T\sim 0.7$--0.8 MK), 19.3 nm, dominated by \ion{Fe}{12} 
($T\sim 1.5$--1.6 MK), and 21.1 nm, dominated by \ion{Fe}{14} 
($T\sim 1.9$--2.0 MK).  The AIA and HMI images were coaligned using 
the IDL procedures {\it read{\_}sdo} and {\it aia{\_}prep} from 
the SolarSoft library.

\subsection{NOAA 12443: 2015 November 5}
As our first illustrative example, Figure 1 shows an evolving collection 
of AR loops inside NOAA 12243, as they appear in the 17.1 and 19.3 nm 
passbands late on 2015 November 5.  The loops are rooted in purely 
positive-polarity plage according to the magnetograms in the 
rightmost column.  As seen most clearly in the 17.1 nm images, 
small, curved structures with dimensions of a few arcseconds are present at 
and around the footpoints of the longer loops.  The looplike fine structure 
appears to be an integral part of the 17.1 nm moss that overlies the 
plage area.  The intensity of the reticulated moss shows substantial 
spatial variations, being considerably brighter inside the circled area 
and in the foreground of the images than in the intervening region, which 
is partially covered by dark filamentary material.  The footpoint structures 
and the overlying emission are also visible in the 19.3 nm images, 
but are much more diffuse.

The reticulated, spongy morphology of the 17.1 nm moss is partly due to 
the presence of cool chromospheric jets (De Pontieu et al., 1999, 2009), 
which may take the form of ``dynamic fibrils,'' mottles, and/or spicules 
(see, e.g., Hansteen et al., 2006; Tsiropoula and Tziotziou, 2004); 
these inclusions give rise to the dark ``holes'' seen in the moss.  
In some cases, a curved feature appearing at the base of an AR loop may be 
a projection effect caused by the presence of foreground chromospheric 
material.  However, small, bright, curved structures are visible throughout 
the moss, and these combine with the dark inclusions to give the moss 
its reticulated appearance.

\subsection{NOAA 12351: 2015 May 23}
The small 17.1 nm loops are often more easily discerned when the moss 
is viewed ``at an angle,'' as when located at relatively high latitudes.  
Figure 2 shows an area of moss inside a sheared, decaying AR (NOAA 12351) 
at latitude N22$^\circ$, as observed on 2015 May 23 in \ion{Fe}{9} 
(left column) and \ion{Fe}{12} (middle column).  A row of looplike features 
with dimensions of a few arcseconds may be seen, for example, inside 
the large circle in the 17.1 nm image recorded at 14:00 UTC.  Some of 
these curved structures are embedded within the purely negative-polarity 
plage, according to the corresponding HMI magnetogram at the far right.  
However, inside the smaller circle at the southwest corner of the 
same 17.1 nm image, similar looplike features are also present 
above a mixed-polarity area at the edge of the plage.  This tends to 
support the conjecture that the small structures inside the plage 
are indeed loops, but that the magnetograms do not show the 
minority-polarity inclusions when the majority-polarity flux is strong 
and heavily dominant.  We also note that the morphological similarity 
between the 17.1 nm fine structure inside and just outside the plage 
suggests that moss is not confined to plages (see also Figure 10 below).

The small looplike features are also visible in \ion{Fe}{12} 19.3 nm, 
as near the center of the large circles in the images recorded at 
14:00 and 14:30 UTC.  This supports the idea that we are observing 
coronal structures. 

\subsection{NOAA 12764: 2020 June 6}
Figure 3 shows an AR loop system observed at high northern latitudes on 
2020 June 6, just after the start of solar cycle 25.  Both ends of the 
loops are embedded in 17.1 nm moss.  In the original and sharpened 
17.1 nm images, bright, curved features with dimensions of 
$\sim$5$^{\prime\prime}$ are visible at the eastern ends (circled) 
of some of the loops; these features seem to be located in purely 
negative-polarity plage, according to the HMI magnetogram (bottom panel).  
The 21.1 nm image also shows compact brightenings within this plage area, 
although their distribution differs from that of the 17.1 nm structures.

\subsection{NOAA 12823: 2021 May 13--14}
Figure 4 shows the configuration of 17.1 nm loops above a negative-polarity 
plage area, as it evolves over a period of a day.  Again, curved features 
with dimensions of several Megameters are clearly visible at the base 
of the AR loops.  In this case, however, some of the footpoint loops 
have at least one leg located above the area of weaker, mixed-polarity field 
outside the unipolar plage. 

\subsection{NOAA 12783: 2020 November 21}
In Figure 5, the western end of a bundle of AR loops is viewed 
``from above'' (rather than ``from the side'' as in the previous examples).  
The footpoints are located inside an area of negative-polarity plage, 
and they are again accompanied by looplike fine structure and 
compact brightenings.  Most of these small features appear to lie 
entirely within the unipolar plage; however, in the 17.1 nm image 
recorded at 04:14 UTC (bottom left panel), the hook-shaped extension 
of the AR loops ends in the mixed-polarity region just northward 
of the plage, as indicated by the corresponding HMI magnetogram 
(bottom right panel).

Figure 6 shows the western end of another loop system situated just to the 
north of that in Figure 5.  Over a 23 minute period, compact brightenings 
at the base of the loops give rise to a succession of narrow outflow streams 
or jets.

\subsection{NOAA 12786: 2020 December 2}
The sequence of 17.1 nm images in Figure 7 provides another example of 
continual jet-like activity at the footpoints of AR loops rooted inside 
or at the edges of unipolar plage areas.  The collimated outflows 
originate from an ever-changing configuration of compact brightenings, 
some of which are recognizable as small loops.

\subsection{NOAA 12791: 2020 December 10}
The compact brightenings and jet-like outflows in unipolar plage areas may 
also be seen at higher temperatures.  The sequence of \ion{Fe}{12} 19.3 nm 
images in Figure 8 illustrates the activity occurring at this location 
throughout the day.  Again, some of the small footpoint features appear 
to be embedded in flux of a single polarity.

\subsection{NOAA 12794: 2020 December 29}
The transient loops in Figure 9 are connected to a large, negative-polarity 
sunspot located to the west of the field of view.  The number, width, 
and intensity of the collimated outflows vary throughout the day.  
The footpoint areas show complex, looplike fine structure that is clearly 
a constituent of the 17.1 nm moss.  This fine structure is present 
even where the underlying flux is weak and of mixed polarity, although 
most of the salt-and-pepper areas are covered by dark fibrilar material, 
which dominates the northern half of the 17.1 nm images.

\section{Nature of the Footpoint Fine Structure: Ephemeral Regions, Granular Motions, and Moss}
According to Fisk (2005), open flux random walks over the solar surface by 
undergoing interchange reconnection with closed loops, and accumulates 
in areas where the density of loops is lowest.  This prediction was 
seemingly confirmed by Abramenko et al. (2006), who found that the 
rate of ER emergence in MDI magnetograms was a factor of 2 lower 
inside coronal holes than in the quiet Sun, and by Zhang et al. (2006), 
who found (using magnetograms taken at the Big Bear Solar Observatory) that 
it was a factor of 3 lower.  Subsequently, however, Hagenaar et al. (2008) 
showed that this reduction applied to unipolar regions in general, 
not just to coronal holes.

As suggested by the examples of the preceding section, the rate of 
ER emergence inside AR plages is likely to be much greater than indicated 
by the magnetograms.  Moreover, the looplike fine structure seen in the 
AIA images appears to be morphologically more or less the same inside and 
just outside the ``unipolar'' plage areas, as illustrated by Figure 2, 
where it is difficult to identify the boundaries between the plage and 
the surrounding mixed-polarity regions from the 17.1 nm images alone.  
We therefore infer that the rate of ER emergence is similar inside 
and outside of unipolar areas, and given by that measured in quiet 
background regions of the Sun.

The ubiquitous small-scale loops have horizontal extents of 
$\sim$2$^{\prime\prime}$--7$^{\prime\prime}$ and lifetimes of a few minutes, 
which are of the same order as the spatiotemporal scales of the 
solar granulation ($\sim$2$^{\prime\prime}$ and $\sim$5--10 minutes).  
The origin of bipoles with total fluxes $\lesssim$10$^{20}$ Mx, 
including in particular ERs, is often attributed to a poorly understood 
``surface dynamo'' that is separate from the main solar dynamo associated 
with sunspots and ARs.  Harvey et al. (2007; see also Lites et al., 1996) 
have identified a patchy, nearly horizontal component of the 
photospheric field with scales ranging from their resolution limit 
of a few arcseconds to $\sim$15$^{\prime\prime}$.  Although they did not 
observe these ``seething'' horizontal fields inside ARs and in the 
magnetic network, we presume that they are also present there but too weak 
to be detected, just as minority-polarity flux is underdetected in 
these regions.  As for their origin, one possibility is that they are 
the lingering, widely dispersed remnants of AR fields that have undergone 
only partial flux cancellation and have not yet fully resubmerged.  
As shown in Wang and Berger (2018), the component of the field parallel to 
the polarity inversion lines of ARs survives many rotations longer 
than the transverse component.  These ``axial'' fields, which are 
closely associated with the helicity of ARs, are only canceled/resubmerged 
after the photospheric neutral lines have themselves been randomized 
on small scales through flux transport processes (see Figure 7 in 
Wang and Berger 2018).\footnote{The axisymmetric (longitude-independent) 
component of the photospheric field, which is the source of the 
polar fields, is also extremely long-lived because it is unaffected by 
differential rotation and decays on the supergranular diffusion timescale.  
However, this component would not contribute to the relatively small-scale, 
horizontal background fields in question.}

The pervasive presence of horizontal fields at or just below the 
solar surface would provide a source for the small loops and ERs 
seen in the EUV images.  These fields would be continually churned 
and twisted by the granular convection, and the twisted small-scale fields 
would continually transfer their energy to the larger coronal loops.  
As in most models for coronal heating, granular motions are the 
basic energy source, but instead of acting directly on the AR loops 
to generate Alfv{\'e}n waves or tangential discontinuities, 
the interaction is mediated through the ubiquitous small-scale fields.

In the standard interpretation, moss represents the transition region 
of hot AR loops (Berger et al., 1999).  However, as already noted, 
comparison of \ion{Fe}{9} 17.1 nm images with magnetograms 
suggests that moss is not confined to strong plages (see also 
De Pontieu et al. (1999, 2003), who found that the distribution of moss 
is poorly correlated with the \ion{Ca}{2} K brightness, a proxy for the 
photospheric field strength).  Figure 10 shows another example of moss-like 
structure overlying a region of relatively weak, mixed-polarity field.  
In our interpretation, moss consists mainly of small loops that are heated 
as their magnetic energy is dissipated by interactions with the 
overlying loops.  However, most quiet Sun areas are covered by 
low-lying 17.1 nm fibrils, which form dark canopies surrounding ARs; 
these ``circumfacular'' regions have long been observed at 
visible wavelengths such as \ion{Ca}{2} K (Hale and Ellerman, 1903; 
St. John, 1911) and H$\alpha$ (e.g., Howard and Harvey, 1964).  As suggested 
in Wang et al. (2011), the dark fibrils tend to form where the 
photospheric field is weak and of mixed polarity and the overlying 
large-scale coronal field has a horizontal orientation, as at the 
edges of flux concentrations; in this case, reconnection produces 
cool, dense, low-lying loops which may eventually evolve into filaments.

\section{Estimating the Contribution of ERs to Coronal Heating in ARs}
Following a procedure similar to that applied to coronal holes in 
Wang (2020), we now estimate the energy flux associated with the 
emergence of ERs (or the continual churning of horizontal fields) 
inside AR plages.

We take as our starting point the MDI measurements of 
Hagenaar et al. (2008, 2010), who found that the rate of ER emergence 
depended on the flux imbalance parameter 
$\xi\equiv\Phi_{\rm net}/\Phi_{\rm abs}$, where $\Phi_{\rm net}$ 
($\Phi_{\rm abs}$) represents the signed (unsigned) flux summed over 
a 92$\times$92 Mm$^2$ area surrounding the ERs.  Their result may be 
expressed as  
\begin{equation}
E_{\rm ER}(\xi)\simeq (1.075 - 0.793\xi^2)\times 10^{-3}\phantom{.}{\rm Mx}
\phantom{.}{\rm cm}^{-2}\phantom{.}{\rm s}^{-1}.
\end{equation}
The measurements were based on 5 minute averages of line-of-sight 
magnetograms with 2$^{\prime\prime}$ pixels, with the individual ERs 
having total unsigned fluxes $\Phi_{\rm ER}\gtrsim 4\times 10^{18}$ Mx.  

Based on the discussion of the preceding sections, we assume that 
the ER emergence rate is the same in unipolar regions and AR plages 
as in mixed-polarity areas of the quiet Sun, and henceforth set $\xi = 0$.  
In addition, we apply two further corrections.  First, following 
Tran et al. (2005), who cross-correlated MDI data with 
saturation-corrected magnetograms from the Mount Wilson Observatory, 
we scale the measured MDI fluxes upward by a factor of 1.7.  Second, 
we allow for the contribution of ERs with total fluxes down to 
$\Phi_{\rm ER}\sim 1\times 10^{18}$ Mx.

The rate of small-scale flux emergence, in the form of ERs and 
intranetwork flux, is known to exhibit an exponential or power-law 
dependence that continues to increase toward smaller spatiotemporal scales 
(see, e.g., Thornton and Parnell, 2011; Zhou et al., 2013).  Because 
our focus is on reconnection events occurring in the corona, we include 
only the contribution of bipoles whose scale sizes exceed $\sim$2 Mm.  
According to Hagenaar (2001), an ER having a total flux of 
$\Phi_{\rm ER} = 1.13\times 10^{19}$ Mx (or $1.92\times 10^{19}$ Mx 
when corrected for line profile saturation) has an average pole separation 
$d_{\rm ER}$ of $\sim$8.9 Mm.  Assuming that 
$d_{\rm ER}\propto\Phi_{\rm ER}^{1/2}$ (following 
Cranmer and van Ballegooijen, 2010) and taking the vertical scale size 
$h_{\rm ER}$ of an ER to be comparable to its pole separation, 
we may write 
\begin{equation}
h_{\rm ER}\sim 2.0\phantom{.}{\rm Mm}\phantom{.}
\left(\frac{\Phi_{\rm ER}}{1\times 10^{18} \phantom{.}{\rm Mx}}\right)^{1/2}.
\end{equation}
This means that ERs with fluxes $\Phi_{\rm ER}\gtrsim 10^{18}$ Mx 
will have loop systems that extend to heights above $\sim$2 Mm, 
the approximate location of the coronal base.

Although they were unable to extend their ER measurements below fluxes 
of $\sim$$4\times 10^{18}$ Mx, Hagenaar et al. (2003) showed that 
the number of emerging ERs as a function of $\Phi_{\rm ER}$ could be 
fitted with an exponential of the form $N_{\rm ER}(\Phi_{\rm ER})\propto
\exp[-\Phi_{\rm ER}/(5\times 10^{18}\phantom{.}{\rm Mx})]$ (see 
their Figure 6).  Scaling the fluxes upward by 1.7, it follows that 
\begin{equation}
E_{\rm ER}(\Phi_{\rm ER}\gtrsim 1\times 10^{18}\phantom{.}{\rm Mx})\simeq 
1.23\phantom{.}E_{\rm ER}(\Phi_{\rm ER}\gtrsim 6.8\times 10^{18}
\phantom{.}{\rm Mx})\simeq 2.25\times 10^{-3}\phantom{.}{\rm Mx}
\phantom{.}{\rm cm}^{-2}\phantom{.}{\rm s}^{-1}. 
\end{equation}

The ERs emerging within a plage will undergo reconnection with the 
overlying AR loops.  According to recent spectropolarimetric analyses 
of plage regions by Morosin et al. (2020), Pietrow et al. (2020), and 
Anan et al. (2021), the strength of the longitudinal field component 
in the chromosphere (at heights of up to $\sim$1 Mm, where the canopy 
has fully formed) is typically on the order of 400 G.  Assuming 
that the field strength at the coronal base, $B_0$, is comparable to 
that in the upper chromosphere, we therefore normalize our estimates 
to $B_0\sim 400$ G.  The recycling time or timescale for an equal amount 
of minority-polarity flux to emerge under the plage field is given by 
\begin{equation}
\tau_{\rm recyc}\sim\frac{B_0}{E_{\rm ER}(\Phi_{\rm ER}\gtrsim 1\times 
10^{18}\phantom{.}{\rm Mx})/2}\sim 99\phantom{.}{\rm hr}\phantom{.}
\left(\frac{B_0}{400\phantom{.}{\rm G}}\right)\left[\frac{2.25\times 10^{-3}
\phantom{.}{\rm Mx}\phantom{.}{\rm cm}^{-2}\phantom{.}{\rm s}^{-1}}
{E_{\rm ER}(\Phi_{\rm ER}\gtrsim 1\times 10^{18}\phantom{.}{\rm Mx})}\right].
\end{equation}
Over this flux replacement time, all of the AR flux will have undergone 
reconnection with the ERs, with magnetic energy being dissipated in 
a layer whose thickness is determined by the heights of the ER loops.  
Denoting the average loop height by $\langle h_{\rm ER}\rangle$, 
we obtain for the energy flux density arising from footpoint reconnection: 
\begin{equation}
F_{\rm ER}\sim\frac{B_0^2}{8\pi}\left(\frac{\langle h_{\rm ER}\rangle}
{\tau_{\rm recyc}}\right)\sim 1.1\times 10^7\phantom{.}{\rm erg}
\phantom{.}{\rm cm}^{-2}\phantom{.}{\rm s}^{-1}\left(\frac{B_0}
{400\phantom{.}{\rm G}}\right)\left(\frac{\langle h_{\rm ER}\rangle}
{6.3\phantom{.}{\rm Mm}}\right)\left[\frac{E_{\rm ER}(\Phi_{\rm ER}\gtrsim 
1\times 10^{18}\phantom{.}{\rm Mx})}{2.25\times 10^{-3}\phantom{.}{\rm Mx}
\phantom{.}{\rm cm}^{-2}\phantom{.}{\rm s}^{-1}}\right].
\end{equation}
Here, the average loop height has been normalized to the value 
corresponding to $\Phi_{\rm ER} = 1\times 10^{19}$ Mx in Equation (2).  
The estimated energy flux density is close to the value of 
$\sim$10$^7$ erg cm$^{-2}$ s$^{-1}$ given by Withbroe and Noyes (1977) 
for the sum of the radiative and conductive energy losses in the 
AR corona.

The above estimate is based on the rate of ER emergence in the 
quiet Sun measured by Hagenaar et al. (2008, 2010).  According to 
Figure 6 of Hagenaar et al. (2008), this rate and thus the recycling time 
are sensitive to the cadence of the magnetograms employed: as the interval 
$\Delta t$ between successive magnetograms decreased from $\sim$1 hr 
to $\sim$5 minutes, $\tau_{\rm recyc}$ decreased from $\sim$13 to 
$\sim$1.5 hr.  This raises the possibility that $E_{\rm ER}$ and thus 
$F_{\rm ER}$ would have been even larger had Hagenaar et al. used 
magnetograms taken less than 5 minutes apart.  In the preceding section, 
we conjectured that ERs may have their origin in the churning of 
ubiquitous horizontal background fields by the solar granulation; 
if so, $\Delta t\sim 5$ minutes might turn out to be the appropriate 
timescale.  However, this remains an important question that needs 
to be addressed by additional measurements.

The energy released in the reconnection events will be in the form of 
ohmic heating, jets, and MHD waves.  The heat deposited near the 
coronal base is conducted both upward into the corona and 
downward through the transition region, with the resulting chromospheric 
evaporation acting to fill the AR loop with hot, dense material.  
We have here shown many examples of upflows, jets, and/or 
outward-propagating intensity fronts that originate from compact 
brightenings at the footpoints of AR loops.  Winebarger et al. (2001) 
identified such upflows in TRACE 17.1 nm images of a bundle of loops 
in NOAA 8395, measuring projected speeds of 5--17 km s$^{-1}$ for the 
intensity fronts.  They conjectured that the flows were related to 
reconnection events, but found no evidence of flux cancellation in 
MDI magnetograms, presumably for the same reason(s) that the HMI magnetograms 
show no underlying minority-polarity flux in many of our examples.

As discussed by Warren et al. (2002, 2003), Winebarger et al. (2003), and 
Aschwanden et al. (2007), impulsive heating at the footpoints of AR loops 
also explains their orders-of-magnitude overdensity compared to that 
predicted by static loop models, the absence of significant temperature 
variations along the lengths of coronal loops (Lenz et al., 1999), 
as well as observations showing that loops initially appear at 
high temperatures before cooling to successively lower temperatures 
(see, e.g., Ugarte-Urra et al., 2006, 2009).  The finding that the 
measured lifetimes of the loops are generally longer than their 
cooling timescales is consistent with AIA observations suggesting 
that a given AR loop may undergo a succession of interactions with 
the underlying small-scale features over periods of up to an hour or more 
(see, e.g., Figures 7--9).

Figure 1 in Wang et al. (2019) illustrates the evolution of an inverted-Y 
structure as observed in \ion{Fe}{9} 17.1 nm and \ion{Fe}{14} 21.1 nm 
during 2013 April 30.  The structure is rooted in a purely unipolar 
(according to the HMI magnetograms) plage area inside NOAA 11731.  
The outflow from the brightened, triangular base area is first seen 
in \ion{Fe}{14}, and then in \ion{Fe}{9} after a lag of $\sim$10 minutes.  
Maintaining this time lag, both outflows reach their maximum intensity 
after $\sim$10 minutes and then fade more gradually over the next 
$\sim$20 minutes.  Warren et al. (2002, 2003) modeled the evolution 
of TRACE loops observed simultaneously in the 17.1, 19.5, and 
28.4 nm filters as involving multiple isothermal threads, with the 
threads having undergone impulsive footpoint heating at different times 
and being in different stages of cooling at any given time.  
Such a multi-thread model would be consistent with the complex topology 
expected of the small-scale footpoint fields as they are continually churned 
by granular motions and undergo reconnection with the overlying large-scale 
coronal field.  The outflow example just described would suggest a 
well-defined onset time for the reconnection (perhaps associated with 
the emergence of an ER), followed by a succession of smaller events 
over a period of tens of minutes.

Reconnection between the AR loops and ER loops extending to heights 
$\gtrsim$2 Mm will also give rise to MHD waves originating near the 
coronal base.  In the model of van Ballegooijen et al. (2011), 
a small fraction of the Alfv{\'e}n waves generated by small-scale 
photospheric convective motions leaks into the corona; the waves 
are trapped inside the loop due to the steep gradients in the 
transition region at each end, interact with each other, and 
undergo turbulent dissipation.  A similar scenario may apply 
if the Alfv{\'e}n waves are generated by reconnection near the 
coronal base.  Because the Alfv{\'e}n crossing times in the corona 
are much shorter than the granular convection timescales on which 
the reconnection is driven, resonant interactions (not considered by 
van Ballegooijen et al.) may also occur between the trapped waves.  
However, in contrast to most models that invoke wave dissipation or 
nanoflares/braiding, we are here in a regime involving strongly antiparallel 
fields.  We would thus expect the main energy release to be in the form 
of ohmic heating at the reconnection site, followed by heat conduction 
upward and downward along the coronal loop and the evaporation of 
material from the top of the chromosphere, as in solar flares.

\section{Summary and Discussion}
Although sometimes still advocated (see, e.g., Priest et al., 2018; 
Syntelis and Priest, 2020), reconnection with small-scale fields or 
the ``magnetic carpet'' appears to have fallen out of favor as a major 
coronal heating mechanism.  Perhaps the main reason for this neglect 
is the near-absence of minority-polarity flux inside AR plages and 
strong network, according to even high-resolution magnetograms.  
This study provides additional support for our earlier conclusion 
that present-day magnetograms greatly underestimate the amount of 
minority-polarity flux at the footpoints of AR loops (Wang, 2016; 
Wang et al., 2019), and suggests that the reconnection rate 
is sufficient to heat the AR corona.  The failure to detect the 
minority-polarity flux may indicate a problem with instrument sensitivity 
when one polarity is heavily dominant and the minority-polarity signal 
is close to the nominal $\sim$10 G noise level.

One of the important points made here (see, e.g., Figures 2 and 10) 
and in our previous studies is that the spatial distribution and 
morphological appearance of looplike fine structure in 17.1 and 19.3 nm 
images seem to be the same inside and just outside ``unipolar'' regions.  
This absence of a well defined boundary for the ``moss'' leads us 
to infer that the rate of small-scale flux emergence (or churning 
of small-scale fields by granular convection) is similar everywhere 
on the Sun, irrespective of the local degree of unipolarity.

Taking the emergence rate to be the same as that measured by 
Hagenaar et al. (2008, 2010) for ERs outside of unipolar areas, 
we have estimated the energy flux density due to reconnection 
with plage fields, $F_{\rm ER}$, to be on the order of 
10$^7$ erg cm$^{-2}$ s$^{-1}$, enough to account for the heating 
of the AR corona.  In general, $F_{\rm ER}\propto 
(B_0^2/8\pi)/\tau_{\rm recyc}\propto B_0$, where $B_0$ is the 
field strength at the coronal base and $\tau_{\rm recyc}\propto B_0$ 
is the timescale for all of the flux to undergo reconnection with 
ERs.  Inside coronal holes, where $B_0\sim 10$ G instead of 
$\sim$300--400 G, $F_{\rm ER}\sim 3\times 10^5$ erg cm$^{-2}$ s$^{-1}$, 
which is sufficient to heat the hole and drive the solar wind.  Using 
energy conservation along a flux tube and tracing from Earth back to 
the solar surface, we have verified that the observed solar-wind 
energy flux density (which is dominated at 1 au by the kinetic energy 
of the bulk flow, $\rho v^3/2$) is indeed proportional to the 
field strength at the coronal base (Wang, 2010, 2020).

Although we have been concerned here only with the heating of the 
corona, it may be that mixed-polarity flux on even smaller scales plays 
an important part in chromospheric heating, given the power law distribution 
of flux emergence.  In their high-resolution spectropolarimetric study 
of chromospheric heating in a plage region, Anan et al. (2021) found 
that the \ion{Mg}{2} radiative flux (a measure of the heating rate) 
was highest in between or at the edges of the individual strong field 
patches within the plage, and was poorly correlated with the local 
chromospheric or photospheric field strength.  Reconnection with 
small bipoles scattered through the plage might account for such 
off-center heating.

Our key prediction is that, rather than simply representing the 
transition region of hot AR loops, ``moss'' actually contains small loops 
that undergo continual interactions with the overlying coronal loops 
and are responsible for their heating.  The {\it Extreme Ultraviolet Imager} 
and the {\it Polarimetric and Helioseismic Imager} on {\it Solar Orbiter} 
may soon provide an opportunity to test this prediction.  In addition, 
we anticipate that the {\it Daniel K. Inouye Solar Telescope}'s 
spectropolarimeters, with their unprecedented sensitivity and 
spatial resolution, will help us to understand better the nature of 
the plage fields.

We thank J. M. Laming for discussions.  This work was supported by NASA 
and the Office of Naval Research.

\clearpage
\begin{figure*}
\vspace{-4.0cm}
\centerline{\includegraphics[width=40pc]{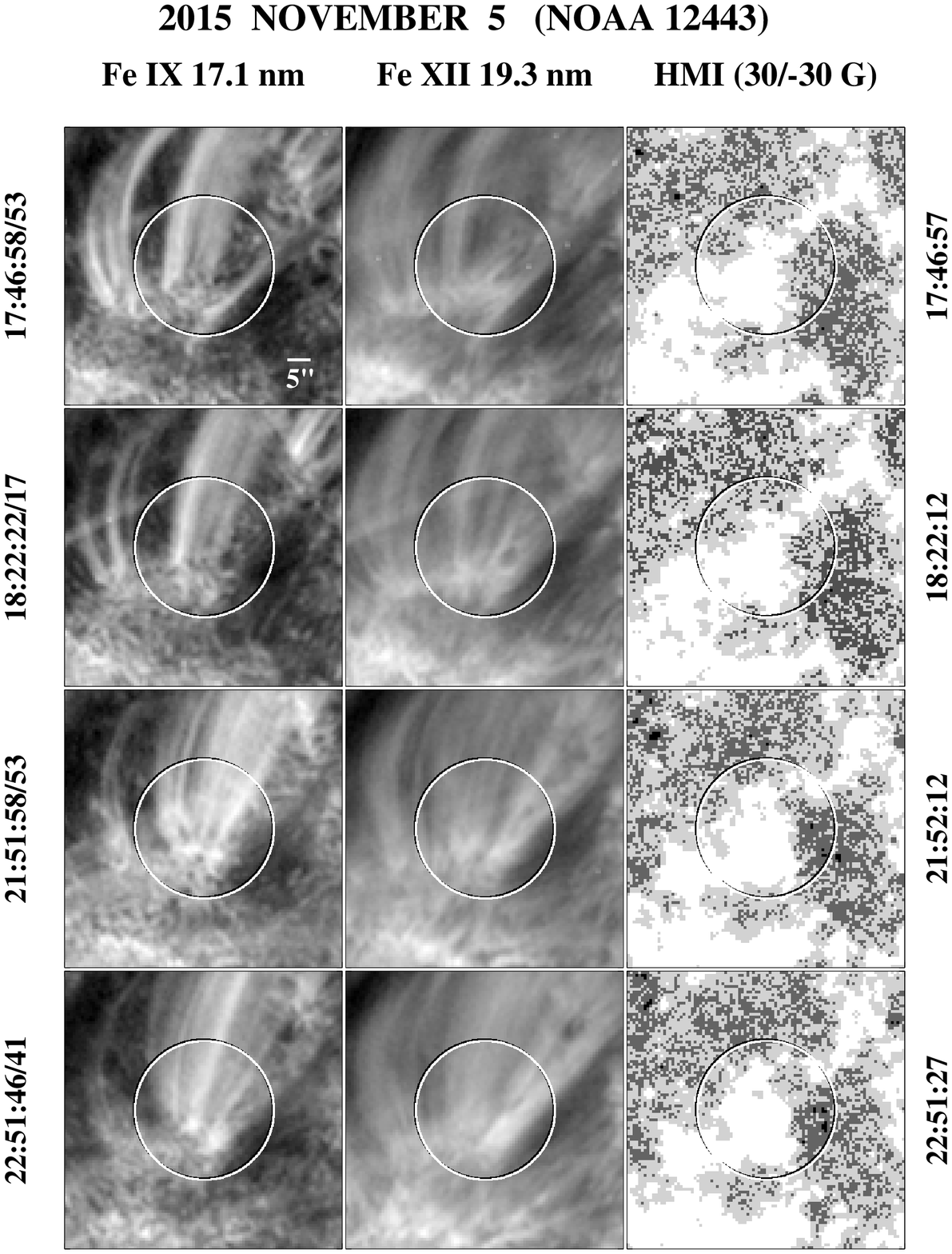}}
\vspace{-2.8cm}
\caption{A cluster of coronal loops observed inside NOAA 12443 late on 
2015 November 5.  Circled area has radius 15$^{\prime\prime}$.  
The Fe IX 17.1 nm (left column) and Fe XII 19.3 nm (middle column) loops 
are rooted in moss overlying a plage area containing only positive-polarity 
flux, according to the corresponding HMI magnetograms (right column).  
The brightness and spatial distribution of the 17.1 nm loops vary 
continually over the 5 hr period, as does the underlying fine structure, 
which, in addition to dark chromospheric material, consists of 
looplike features and compact brightenings that appear to be 
an integral part of the moss.  The emission is more diffuse and 
widespread in the 19.3 nm images (recorded 5 s before the 17.1 nm images), 
but the brightenings are still visible at the loop footpoints.  Movies 
suggest that the fading of the \ion{Fe}{9} loops is often accompanied 
by downflows (see the animation accompanying Figure 10 in Wang et al., 2019).  
Here and in subsequent figures, the gray scale used in the 
magnetograms is as follows: white ($B_{\rm los} > 30$ G); 
light gray (0 G $< B_{\rm los} < 30$ G); 
dark gray ($-$30 G $< B_{\rm los} < 0$ G); black ($B_{\rm los} < -30$ G).}
\end{figure*}

\begin{figure*}
\vspace{-4.1cm}
\centerline{\includegraphics[width=42pc]{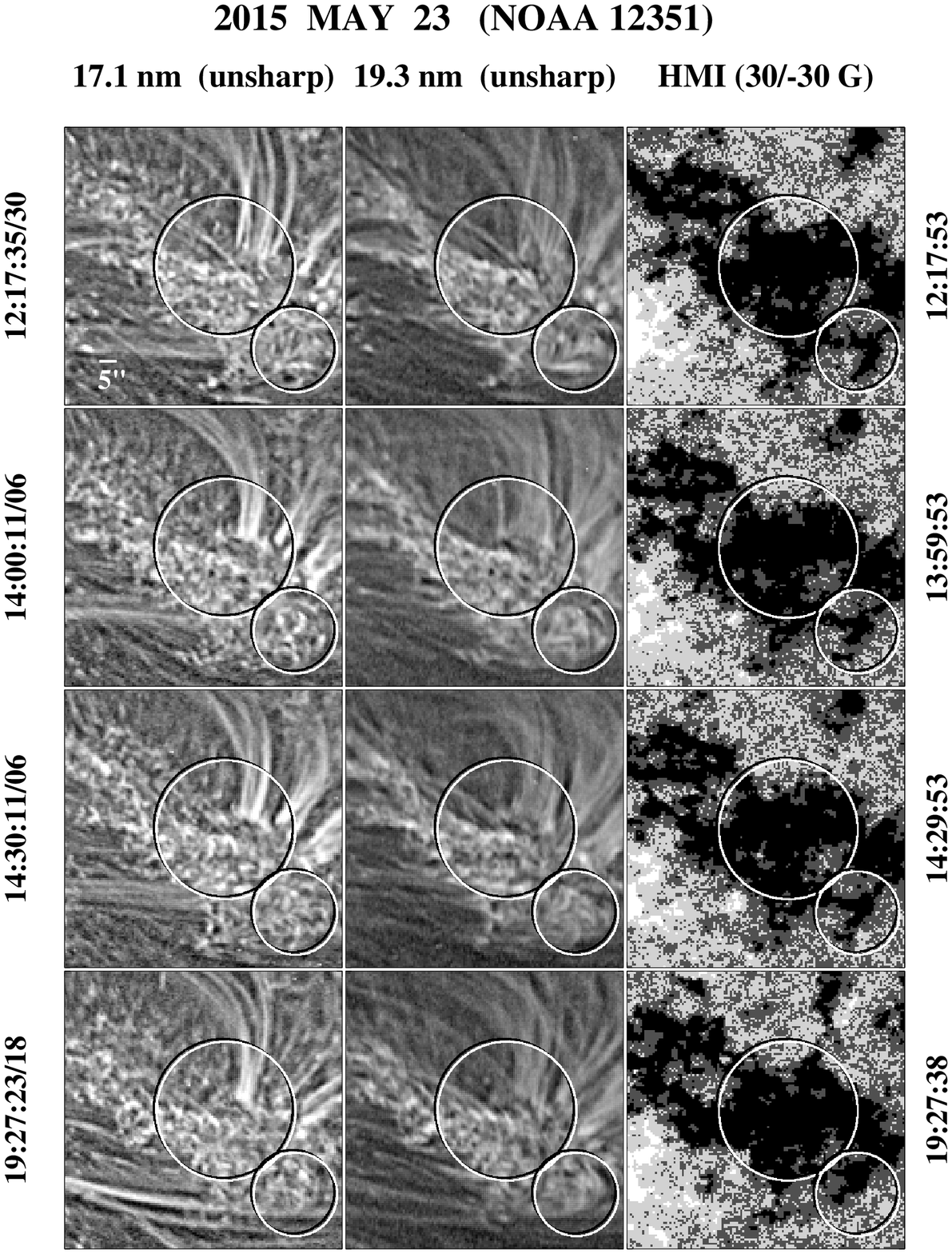}}
\vspace{-2.8cm}
\caption{Small loop-shaped features embedded within the moss/plage inside 
NOAA 12351, as observed on 2015 May 23.  Left column: Fe IX 17.1 nm 
images recorded at 12:17:35, 14:00:11, 14:30:11, and 19:27:23 UTC.  
Middle column: Fe X11 19.3 nm images recorded at 12:17:30, 14:00:06, 
14:30:06, and 19:27:18 UTC.  Right column: line-of-sight HMI magnetograms 
recorded at 12:17:53, 13:59:53, 14:29:53, and 19:27:38 UTC and 
saturated at $\pm$30 G.  An unsharp mask has been applied to the EUV images.  
Note that the moss-like structure and looplike features extend into 
the mixed-polarity, weaker-field areas beyond the plage itself, as indicated 
for example by the circled region at the southwest corner of the images.  
Loop-shaped structures with dimensions of a few arcseconds may be 
discerned not only in Fe IX but also in Fe XII, as near the center of 
the large circles in the 19.3 nm images taken at 14:00 and 14:30 UTC.}
\end{figure*}

\begin{figure*}
\vspace{-5.5cm}
\centerline{\includegraphics[width=45pc]{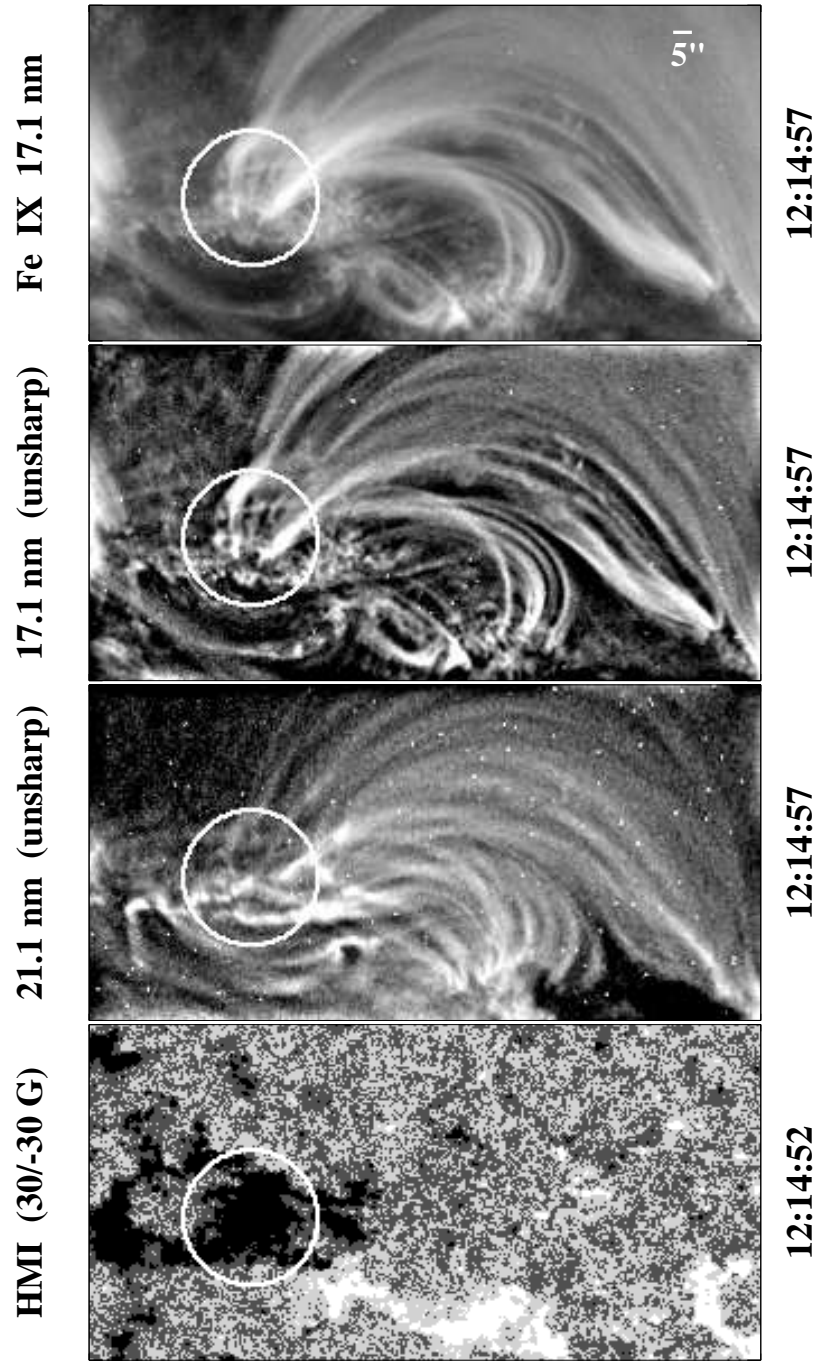}}
\vspace{-2.7cm}
\caption{AR loop system (NOAA 12764) observed on 2020 June 6, at the onset 
of solar cycle 25.  Within the circled area (of radius 18$^{\prime\prime}$), 
looplike features with dimensions on the order of 5$^{\prime\prime}$ 
may be seen at the footpoints of the longer loops.  These features are 
embedded in the 17.1 nm moss overlying the negative-polarity plage.}
\end{figure*}

\begin{figure*}
\vspace{-3.8cm}
\centerline{\includegraphics[width=42pc]{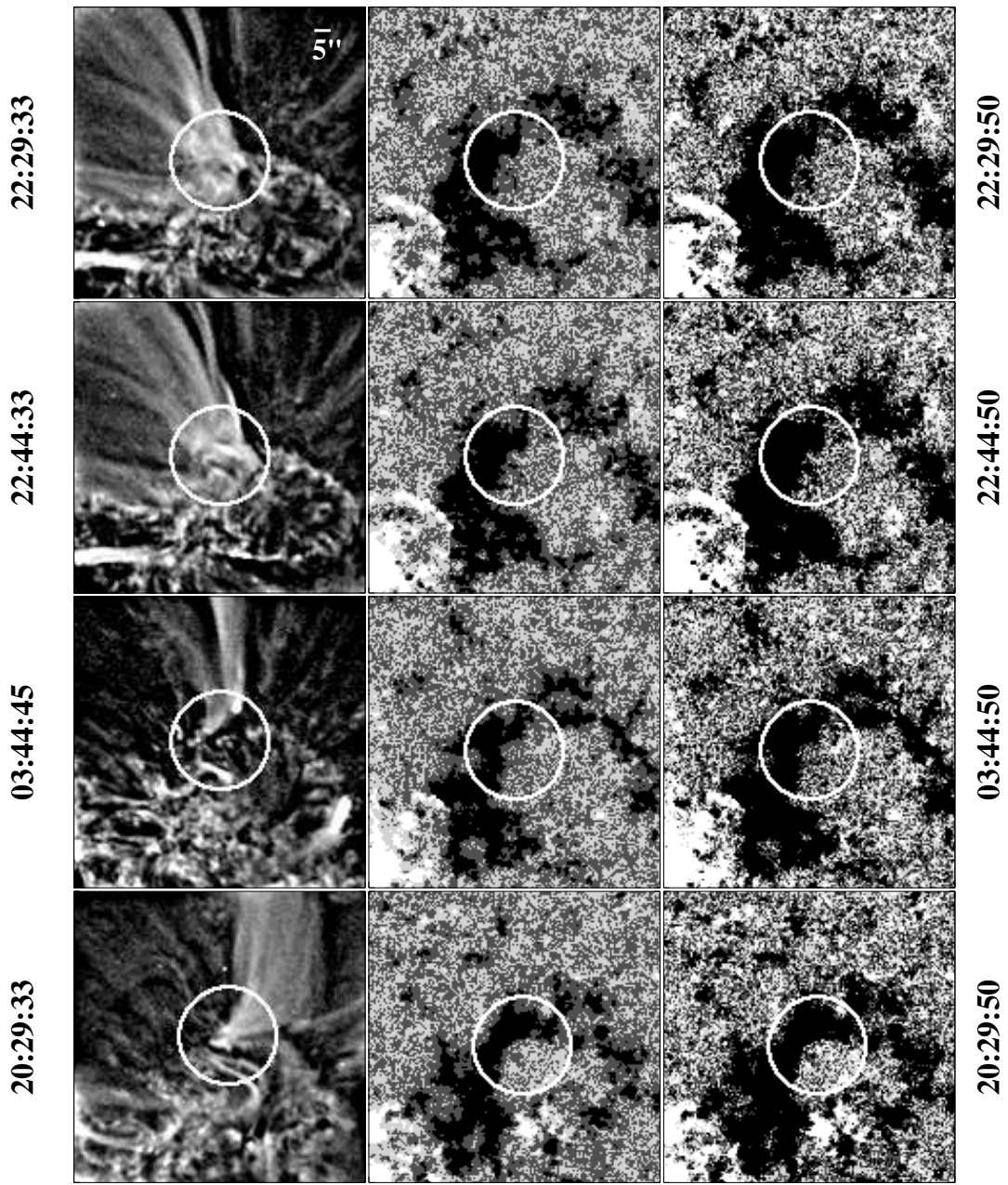}}
\vspace{-2.5cm}
\caption{Evolution of 17.1 nm loops inside NOAA 12823, 2021 May 13--14.  
As the underlying plage area narrows, the cluster of AR loops present 
late on May 13 fades and is replaced by a smaller bundle on May 14.  
Relatively large ($\gtrsim$5$^{\prime\prime}$) looplike features may be 
seen at the base of the AR loops; some of these footpoint structures 
have at least one leg located above the weak, mixed-polarity region 
on the west side of the negative-polarity plage.  The magnetograms 
in the rightmost column have been saturated at $\pm$10 G instead of $\pm$30 G 
(as in the middle column), so that light and dark gray pixels here indicate 
locations where $B_{\rm los}$ is close to the HMI noise level.}
\end{figure*}

\begin{figure*}
\vspace{-5.0cm}
\centerline{\includegraphics[width=42pc]{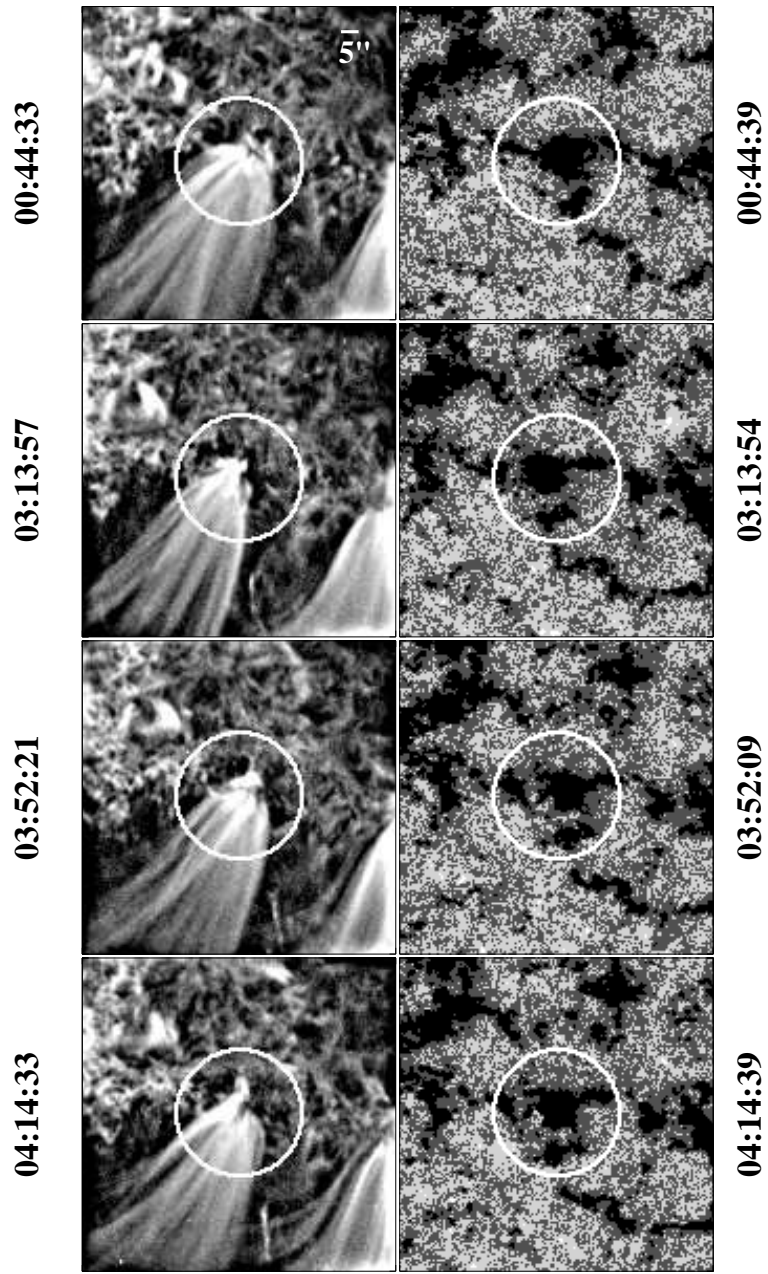}}
\vspace{-2.5cm}
\caption{The western end of a bundle of 17.1 nm loops is here viewed 
``from above.''  The loops are rooted in a fragment of negative-polarity 
plage on the west side of NOAA 12783, 2020 November 21.  Small, bright 
features may be seen at the terminus of the loops, some of which 
are embedded within the unipolar plage.}
\end{figure*}

\begin{figure*}
\vspace{-5.0cm}
\centerline{\includegraphics[width=42pc]{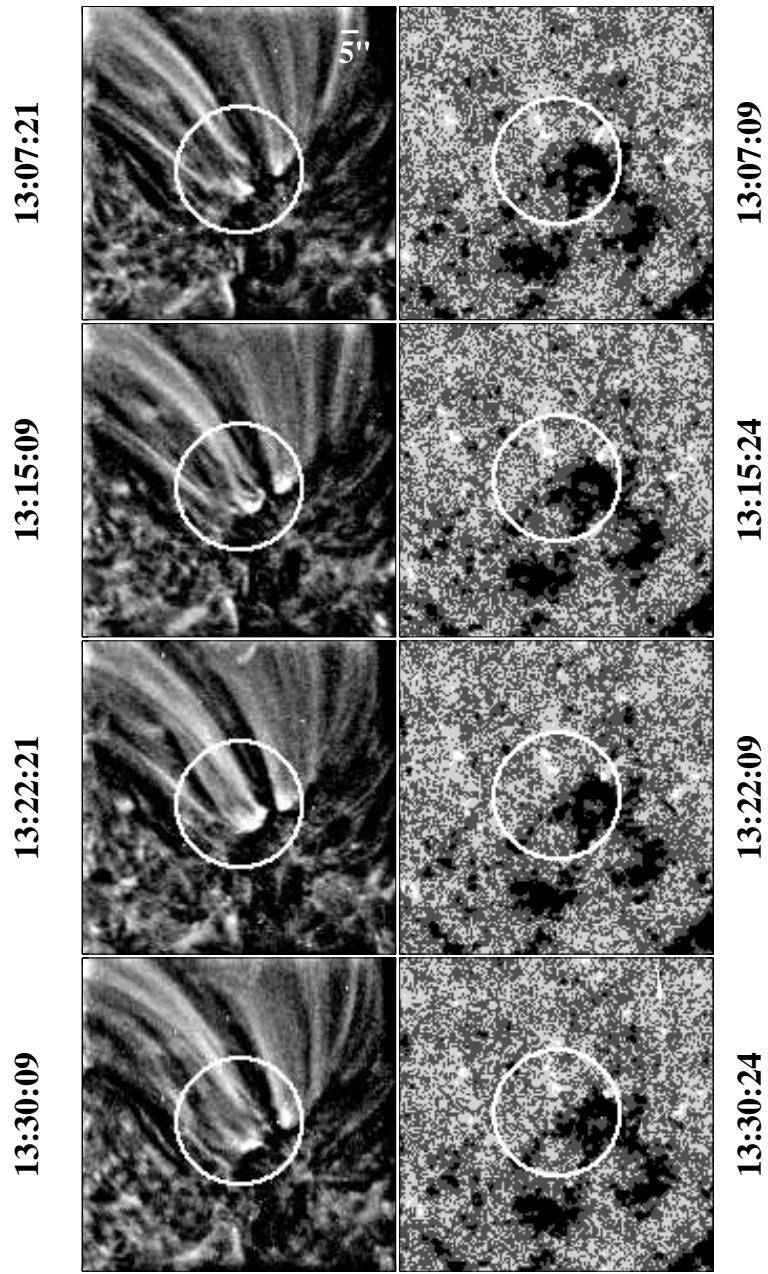}}
\vspace{-2.5cm}
\caption{Compact brightenings at the footpoints of 17.1 nm loops, 
as observed near the northwestern edge of NOAA 12783 on 2020 November 21.  
The intensities and spatial distribution of the footpoint brightenings, 
some of which are accompanied by jets or narrow outflows along the 
overlying loops, vary continually on timescales of minutes.}
\end{figure*}

\begin{figure*}
\vspace{-5.0cm}
\centerline{\includegraphics[width=42pc]{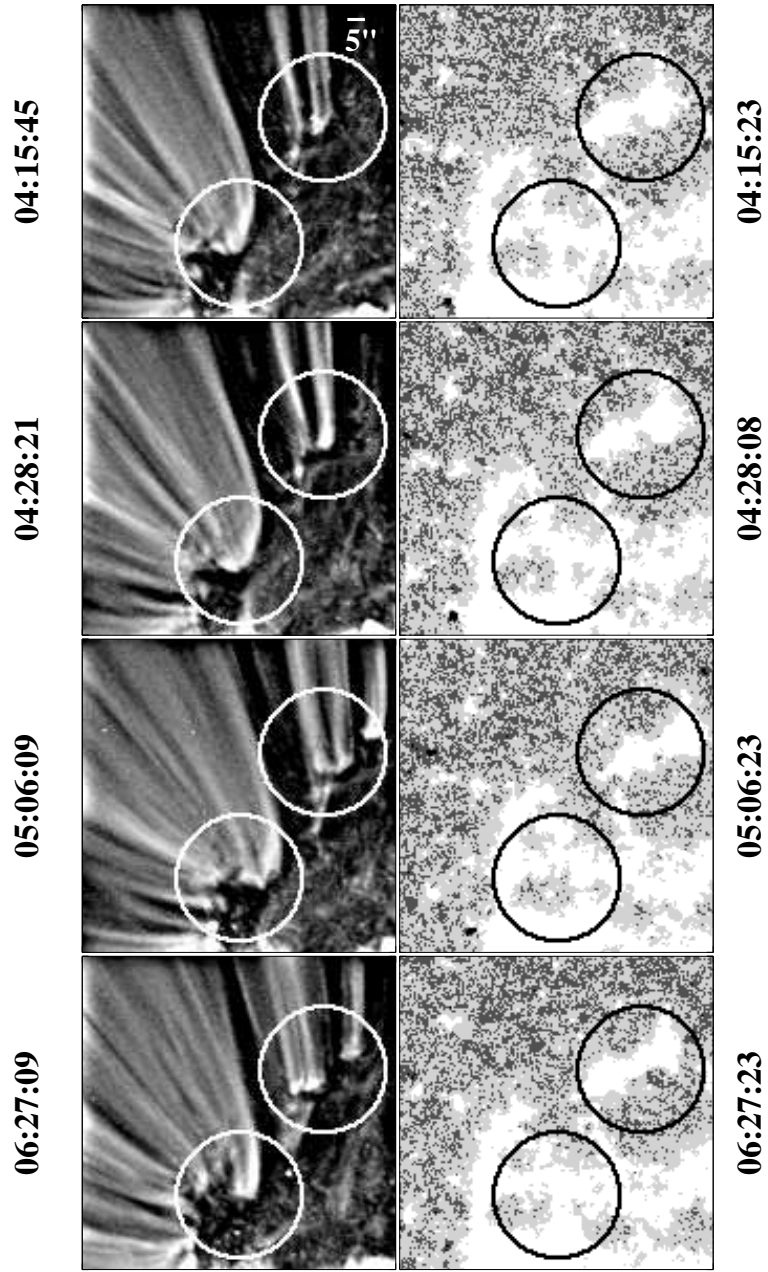}}
\vspace{-2.5cm}
\caption{More examples of 17.1 nm AR loops with compact brightenings at 
their bases.  The loops in this case are rooted in and around a unipolar 
plage area inside NOAA 12786.  Again, bright jet-like outflows may be seen 
emanating from many of the footpoint features.}
\end{figure*}

\begin{figure*}
\vspace{-5.0cm}
\centerline{\includegraphics[width=42pc]{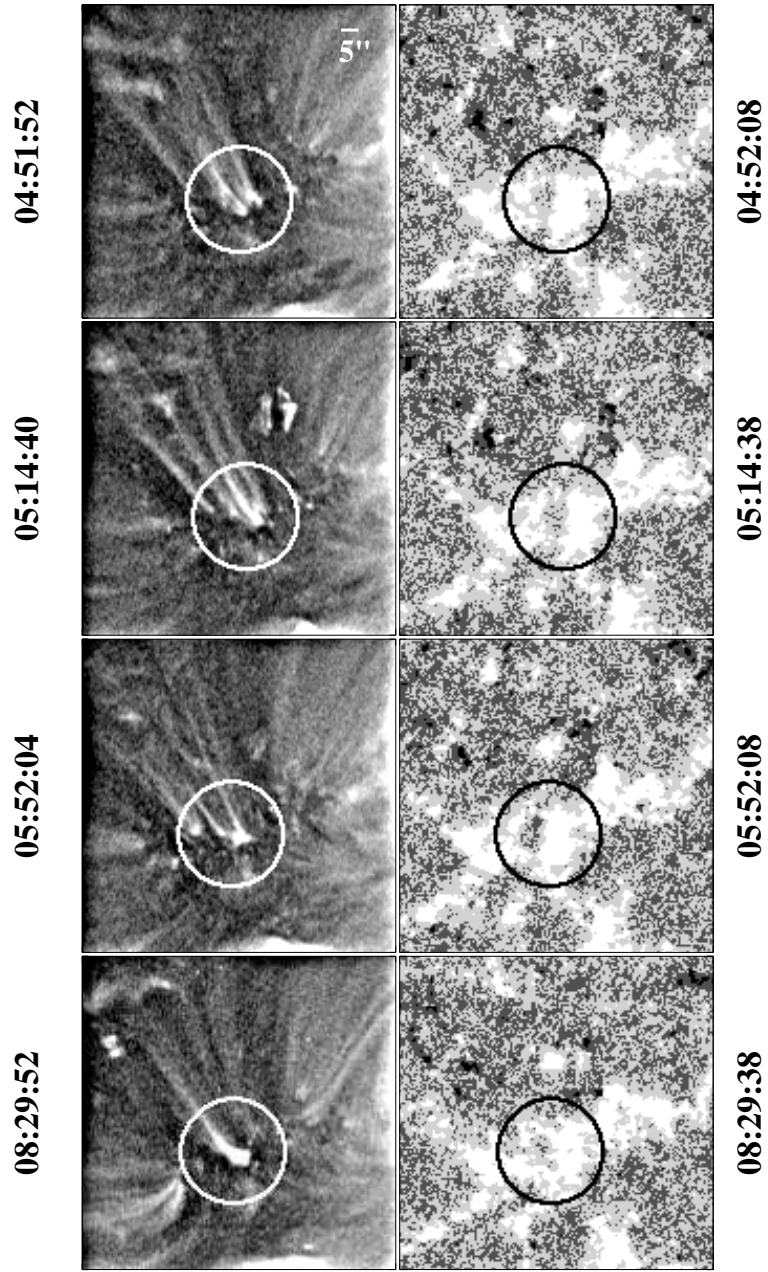}}
\vspace{-2.5cm}
\caption{Compact brightenings at the footpoints of Fe XII 19.3 nm loops 
in NOAA 12791, 2020 December 10.  Again, the footpoint brightenings 
are accompanied by jet-like outflows, and some of the compact features 
appear to be embedded in purely unipolar flux.}
\end{figure*}

\begin{figure*}
\vspace{-5.0cm}
\centerline{\includegraphics[width=42pc]{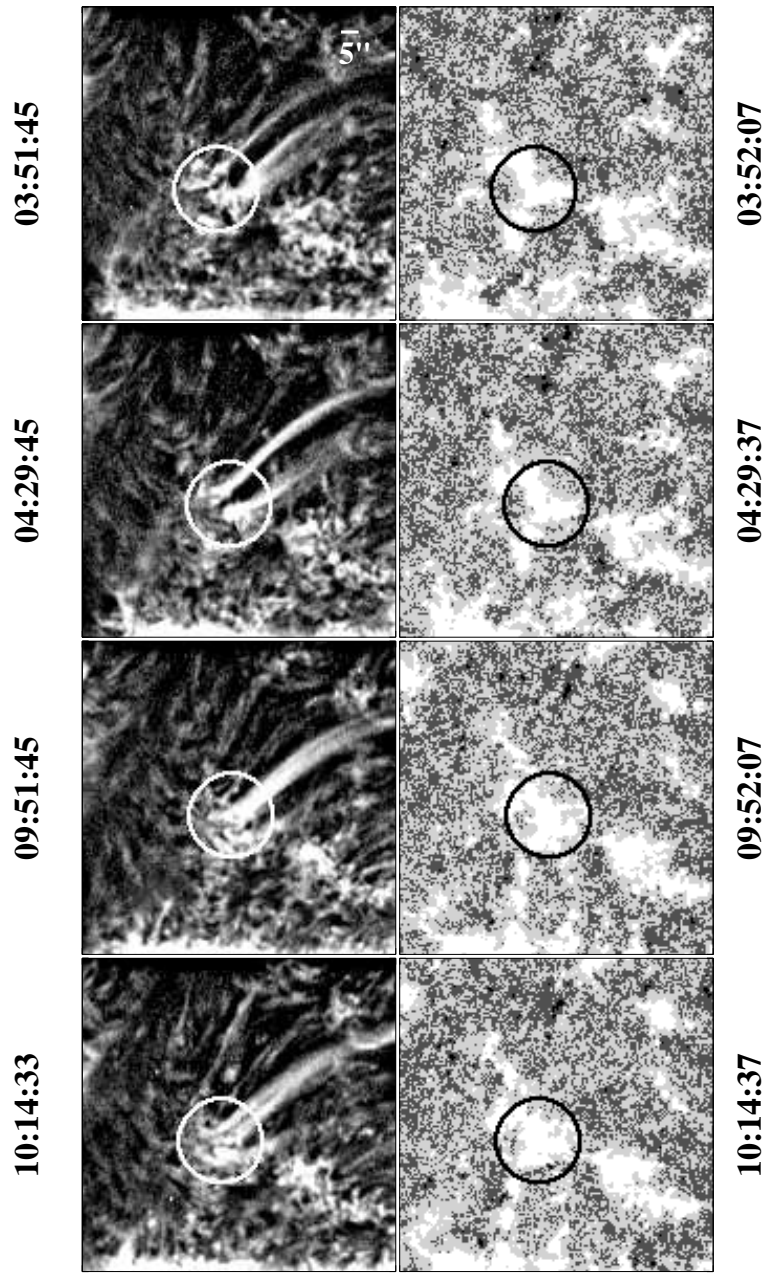}}
\vspace{-2.5cm}
\caption{The highly time-varying 17.1 nm outflows seen here are directed 
toward a large, negative-polarity sunspot in NOAA 12794.  Note the 
complex fine structure at the bases of the loops.}
\end{figure*}

\begin{figure*}
\vspace{-4.0cm}
\centerline{\includegraphics[width=45pc]{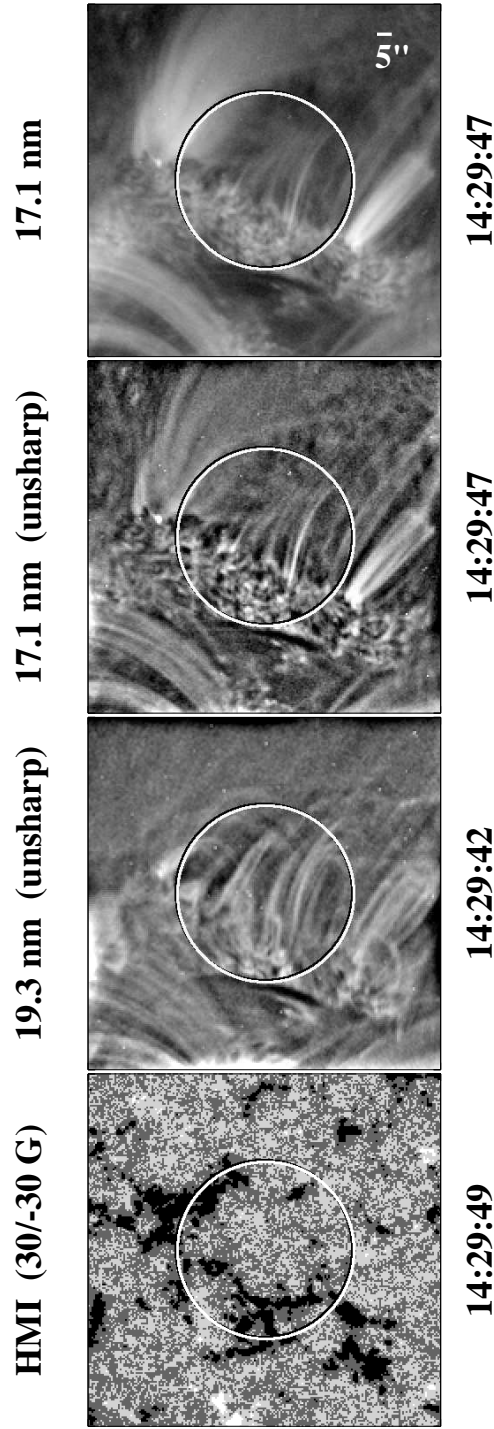}}
\vspace{-3.0cm}
\caption{An example of 17.1 nm ``moss'' overlying an area of relatively 
weak photospheric field.  The obvious presence of mixed-polarity flux 
in this case supports the idea that small loops are a major constituent 
of moss, including that observed in ``unipolar'' plage areas.  
(See also Figure 2, where the small circles highlight moss- and 
looplike structure extending beyond the edge of the plage.)}
\end{figure*}

\end{document}